\documentclass[11 pt]{amsart}
\usepackage[colorlinks=true, pdfstartview=FitV, linkcolor=blue, citecolor=blue, urlcolor=blue]{hyperref} 
\usepackage{graphicx, multicol}
\usepackage{xcolor}
\begin{document}

\thispagestyle{empty}

\begin{center}{\large \bf  \parbox{6.2in}{
Misleading assertions, unjustified assumptions, and additional limitations of a study by Patone et al., described in the article ``Risk of Myocarditis After Sequential Doses of COVID-19 Vaccine and SARS-CoV-2 Infection by Age and Sex'' }}
\smallskip

Paul Bourdon\footnote{Professor of Mathematics, General Faculty, at the University of Virginia (Retired); Formerly, Cincinnati Professor of Mathematics at Washington \& Lee University},  PhD; Spiro Pantazatos\footnote{Assistant Professor of Clinical Neurobiology at Columbia University Irving Medical Center; Molecular Imaging and Neuropathology Division, New York State Psychiatric Institute, New York, NY\\[1pt] \rule{.175in}{0in} 2010 {\it Mathematics Subject Classification}: Primary 92C50; Secondary 37M99}, PhD 
\end{center}

\bigskip
{\small 
{\sc Abstract}:   We describe several shortcomings of a study by Patone et al., whose findings were recently published in the American Heart Association Journal {\it Circulation}, including the following:
  \begin{itemize}
\item The study's principal conclusion, as initially stated, is ``Overall, the risk of myocarditis is greater after SARS-CoV-2 infection than after COVID-19 vaccination and remains modest after sequential doses including a booster dose of BNT162b2 mRNA vaccine.'' However, Patone et al.\ never attempt to assess the incidence of myocarditis in their study population following SARS-CoV-2 infection. Rather, they make an untenable assumption that all infections occurring in their study population are associated with (reported) positive COVID-19 tests. Using publicly available data from the UK's ONS and NHS, we show that Patone et al.'s estimates, for the unvaccinated, of myocarditis incidence associated with {\it infection} are likely overestimated by a factor of at least 1.58.  When this factor is taken into account, in, e.g., the incidence rate ratios (IRRs) of the authors' Table 3, we find that, for men under age 40, the risk of myocarditis after dose 2 of Pfizer's BNT162b2 is higher than post-infection risk in the unvaccinated (while Table 3 suggests the opposite is true).
\item  The method Patone et al.\ use to compute the incidence of myocarditis among the unvaccinated after a positive COVID test may overestimate risk. The authors assume, without justification, that unvaccinated persons hospitalized during the study period with positive-test-associated myocarditis would later choose to vaccinate (and thereby enter the study population already having experienced the event of interest) with the same probability as unvaccinated persons who have had a positive SARS-CoV-2 test.   We present a plausibility argument that suggests a possible further exaggeration of myocarditis hospitalization risk post infection by a factor of 1.5.
\item Patone et al.\ fail to discuss important limitations of their study with respect to guiding public health recommendations.  For instance, the study period is 1 December 2020 until 15 December 2021.  An insignificant number of cases contributing to the study's findings were Omicron-variant cases.  Thus, the study's estimates of myocarditis risk following infection do not speak to the risk following Omicron infection, which is recognized to be milder than that of previous variants. In fact, a study by Lewnard et al.\  suggests hazard ratios for severe clinical outcomes are reduced across the board for Omicron versus Delta, with hazard reduction ``starkest among individuals not previously vaccinated against COVID-19''; e.g., the adjusted hazard ratio for mortality is 0.14 (0.07, 0.28) for the unvaccinated.  Thus, relative to Omicron, we expect that myocarditis incidence rates following infection will be lower than even the appropriately corrected rates (see the preceding two bullet points) based on Patone et al.'s data.  
\end{itemize}     }

\bigskip

\section{Introduction}
 On 22 August 2022,  \href{https://www.ahajournals.org/doi/abs/10.1161/CIRCULATIONAHA.122.059970}{\textcolor{blue}{\underline{a research article}}} \cite{Patone}, titled ``Risk of Myocarditis After Sequential Doses of COVID-19 Vaccine and SARS-CoV-2 Infection by Age and Sex'' by Patone et al.\ was published in the American Heart Association Journal {\it Circulation}.  The article reports on a case-series study, stratified by age and sex, designed to evaluate the association between COVID-19 vaccination and myocarditis as well as between COVID-19 infection and myocarditis. For Patone et al.'s study, a case of myocarditis is one that results in death or  in  hospital admission for myocarditis---some of these admissions occurred in temporal proximity (1 to 28 days) to a COVID-19 vaccination, some in temporal proximity to a positive COVID test, and some,  ``baseline cases,'' did not have either of these temporal associations.
 
   Patone et al.'s study population consists of 42,842,345 residents of England, ages 13 and up, receiving at least one dose of a COVID-19 vaccine during the study period 1 December 2020 until 15 December 2021.   Over the course of the study period, 5,934,153 (13.9\%) of the study population tested positive for SARS-CoV-2, including 2,958,026 (49.8\%) before their first vaccination.    
  
  \section{Risk of Myocarditis After Infection} 
   The principal conclusion of the {\it Circulation} article by Patone et al.\ is stated on its  first page as follows: 
   \begin{quotation}
   {\small Overall, the risk of myocarditis is greater after SARS-CoV-2 infection than after COVID-19 vaccination and remains modest after sequential doses including a booster dose of BNT162b2 mRNA vaccine. However, the risk of myocarditis after vaccination is higher in younger men, particularly after a second dose of the mRNA-1273 vaccine. }
   \end{quotation}
 On page 2, the authors state their conclusion differently, replacing ``SARS-CoV-2 infection'' with ``positive SARS-CoV-2 test'':
 \begin{quotation}
  {\small It is important that we also demonstrated across the entire vaccinated population in England that the risk of myocarditis after vaccination was small compared with the risk after a positive SARS-CoV-2 test.}  
  \end{quotation}
  
     However, the authors overestimate risk of myocarditis following {\it infection}. In the section ``Exposures'' on page 3,  Patone et al.\ state \begin{quotation} {\small   The exposure variables were a first, second, or booster dose of the ChAdOx1, BNT162b2, or mRNA-1273 vaccines, and SARS-CoV-2 infection, defined as the first SARS-CoV-2– positive test in the study period. } \end{quotation} The authors estimate SARS-CoV-2 infections in the study population using the number of positive SARS-CoV-2 tests recorded for its members but they do not account for underascertainment bias (i.e. infections that occur that are not associated with a positive test.)
     The number of COVID infections in a population may be 3 to 4 times the number of documented COVID cases among its members; for example, an ``Interactive COVID-19 Event Risk \href{https://covid19risk.biosci.gatech.edu}{\textcolor{blue}{\underline{Planning Tool}}},'' developed with support from several U.S.\ sources including the Centers for Disease Control (CDC) and the National Science Foundation, defaults to $3\times$ to obtain an estimated number of infections from number of cases for``global events,'' based on data from a number of countries, including the UK. The  \href{https://www.cdc.gov/coronavirus/2019-ncov/cases-updates/burden.html}{\textcolor{blue}{\underline{CDC estimates}}}  that  $4\times$ is the appropriate multiplier (95\% CI: 3.4, 4.7), at least for the period February 2020--September 2021 in the U.S.

     Also,  in assessing the risk of myocarditis after infection, every infection should be counted, including repeat infections (some of which would be documented by a positive test result).   In short, the authors have defined ``SARS-CoV-2 infection'' in such a way that their study is guaranteed to exaggerate the true risk of myocarditis after infection---for every case of myocarditis in an infected member of the study population, there will be a positive SARS-CoV-2 test (e.g., one administered by the hospital where the member received treatment for myocarditis); however, for many infections in the study population, there will not be a corresponding positive test. In assessing the number of infections in unvaccinated persons who later join the study population through getting an initial dose of a COVID vaccine, we cannot simply multiply a given number of positive-test results by 3 or 4.  The number of unvaccinated who eventually join the study population starts at 42,842,345 and gradually declines---we must keep track of these declining numbers as well time-varying rates of infection.

     We are not the first to notice that Patone et al's study exaggerates the risk of myocarditis after SARS-CoV-2 infection.   For instance,  Dr.\ \href{https://brownstone.org/articles/myocarditis-under-age-40-an-update/}{\textcolor{blue}{\underline{Vinay Prasad raised this issue}}} 28 December 2021 (in commenting on an earlier publication based on study data from the period  1 December 2020 to 24 August 2021):  
   \begin{quotation} {\small  
     While the denominator for vaccines is known with precision, the true number of infections is unknown. Many people don’t seek testing or medical care. So the red bar above [illustrating positive-test-associated excess myocarditis cases]  will be shorter if you used a sero-prevalence (aka the correct) denominator. The authors needed to fix this.}
     \end{quotation}
     Unfortunately,  the authors (Patone et al.) did not ``fix this''  in their recent article appearing in {\it Circulation}.   Thus, we offer a fix here.
     
    As we noted earlier, 2,958,026 of the study population tested positive for SARS-CoV-2 before their first vaccination; 114 myocarditis cases occurred during the study period in this subset of the population 1--28 days from the test date.  Based on this raw data (used in Patone et al's ``Main Analysis''), the incidence of positive-test-associated myocarditis among study-population members while unvaccinated is
\smallskip

\begin{center}
 $ \dfrac{114}{2,958,026} \approx 38.54$ cases per million positive tests per 28 days.\footnote{Remark: using the IRR of 11.14 from the ``Main Analysis'' portion of Table 3, third column from the right, we find the  corresponding baseline incidence is approximately $38.54/11.14 \approx  3.46$ cases per million per 28 days; thus, we arrive at $38.54-3.46 = 35.08\approx 35$ excess cases per million per 28 days, the number reported in Patone et al's Table 4.}
\end{center}
\smallskip
To obtain myocarditis incidence after a COVID {\it infection}, we must increase the denominator in the preceding quotient so that it reflects the number of SARS-CoV-2 infections that occurred  in study-population members while they were unvaccinated. 
      
   Using publicly available  \href{https://www.ons.gov.uk/peoplepopulationandcommunity/healthandsocialcare/conditionsanddiseases/articles/coronaviruscovid19infectionsurveytechnicalarticlecumulativeincidenceofthenumberofpeoplewhohavetestedpositiveforcovid19uk/22april2022#estimates-of-cumulative-incidence-by-country}{\textcolor{blue}{\underline{data}}} \cite{ONSArt} from the UK's Office of National Statistics  estimating cumulative percentages of England's population that would have---assuming universal testing---tested positive for SARS-CoV-2 through a given date as well as National Health Service  \href{https://www.england.nhs.uk/statistics/statistical-work-areas/covid-19-vaccinations/covid-19-vaccinations-archive/}{\textcolor{blue}{\underline{data}}} \cite{NHSCVA} providing the total number of 1st doses of COVID vaccines administered through a given date, we have computed 4,685,095 as a lower bound on the number of SARS-CoV-2 infections occurring during the study period in members of the study population while they were unvaccinated. (See the appended supplementary analysis  ``Estimating the Number of SARS-CoV-2 Infections in Members of Patone et al.'s Study Population Before Vaccination.'')  Thus, an estimate of  the incidence of myocarditis after a COVID {\it infection} among study-population members while unvaccinated  is   
$$
  \dfrac{114}{4,685,095} \approx 24.33\  \text{cases per million infections per 28 days},\footnote{``infection'' here means ``first infection of the study period''}
  $$
  and the preceding is likely to be an overestimate because the method of computing infections, described in the supplement, produces a lower bound on number of infections based on ONS and NHS data.

  To understand the implications of using a more realistic count of SARS-CoV-2 infections occurring among members of the study population before they received an initial dose of a COVID vaccine, we'll assume that the ratio of infections to positive tests, $1.58 \approx 4685095/2958026$,  is similar for the four major demographic groups considered in the study: Men $< 40$, Women $< 40$, Men $\ge 40$, Women $\ge 40$.  Now consider the data in Patone et al's Table 3 that express the risk to men under 40 of contracting myocarditis after COVID vaccination, or a positive SARS-CoV-2 test, in terms of  incident-rate ratios (IRRs):

   \begin{center}
   \includegraphics[height=2.7in]{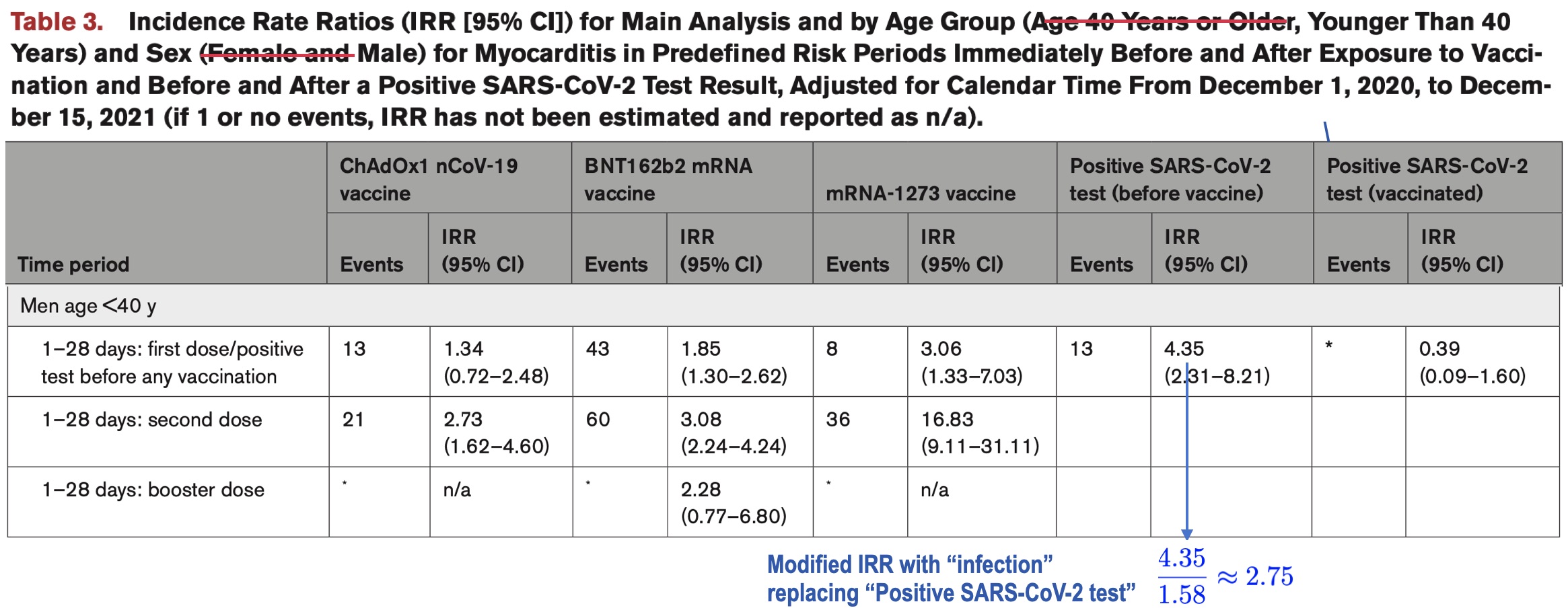}
   \end{center}
   We have modified Table 3 from Patone et al.'s article, eliminating rows corresponding to other demographic groups and adjusting the description of table contents with appropriate strikethroughs.  In the third column from the right, the table has 4.35 as the IRR:  $4.35 = M/B$, where $M$ is the test-positive-associated myocarditis incidence among unvaccinated men under 40 who would later join the study population via vaccination\footnote{$M =13/(\text{number of men under 40 in the study population, in millions, testing positive while unvaccinated})$} and  $B$ is the baseline incidence. We have $M = 4.35 B$, and  because we are assuming the number of infections is roughly 1.58 times the number of  (first) positive tests, the incidence of myocarditis after infection should be approximately $M/1.58$ and the corresponding IRR is $\frac{M/1.58}{B} = \frac{1}{1.58} \frac{M}{B} = \frac{1}{1.58} 4.35 \approx 2.75$.  Note this IRR falls below that for the second dose of Pfizer's BNT162b2, namely 3.08, as well as a first dose of Moderna's mRNA-1273.  
    \section{Unjustified Assumptions}
    The method Patone et al.\ use to compute the incidence of myocarditis among the unvaccinated after a positive COVID test is flawed, and in a way that quite possibly leads to an overestimate of the risk.   In the preceding section, we established that Patone et al.'s denominator in their calculation of incidence of infection-associated myocarditis  among members of the study population while they were unvaccinated is too small, resulting in overstatement of the incidence.  In this section, we show that the numerator, namely 114, in this calculation is problematic as well. 
    
   Recall that  for Patone et al.'s study, a case of myocarditis is one that results in death or  in  hospital admission for myocarditis.    Because the study-population consists of only vaccinated individuals and any unvaccinated person who dies from myocarditis in temporal proximity to a positive COVID test will not be able to later vaccinate, the numerator 114 won't include any cases of myocarditis resulting in death.

    Furthermore, COVID-related myocarditis risk among the unvaccinated is, of course, unrelated to vaccination.  Yet, the study population consists of only vaccinated individuals.  This creates an illogical dependence of Patone et al.’s  computation of the incidence of positive-test-associated myocarditis among the unvaccinated on the decision to later vaccinate or not made by a very small number of individuals in England—those individuals, ages 13 and up, hospitalized with positive-test-associated myocarditis during the study period while unvaccinated.  We know 114 of those individuals later chose to vaccinate, but we do not know how many chose not to vaccinate.  What if none had chosen to vaccinate? Then, the numerator 114 in Patone et al.’s main analysis of incidence would be 0 and the study would have shown 0 risk of myocarditis among the unvaccinated.   We prove below that Patone et al.’s claimed incidence of positive-test-associated hospitalization for myocarditis among the unvaccinated is valid if and only if during the study period  unvaccinated persons (age 13 and up) hospitalized with positive-test-associated myocarditis later chose to vaccinate with the same probability as unvaccinated persons (age 13 and up) who already had a positive SARS-CoV-2 test.

    Let's call a case of myocarditis in a study-population member  for which the associated hospital admission is within 28 days of a positive SARS-CoV-2 test and for which the member is not yet vaccinated, a ``test positive u-myocarditis case''; let's call the subset of of the study population having a positive SARS-CoV-2 test  during the study period while unvaccinated the ``test positive u-subpopulation''.   As we have already noted, based on raw data, the incidence of positive-test-associated hospitalization for myocarditis among members of the study population not yet vaccinated is  
    $$ 
 Q :=   \frac{\text{number of test-positive u-myocarditis cases}}{\text{number of members of the test positive u-subpopulation}} = \frac{114}{2958026} \approx 38.54.
    $$
      However it is important to recognize that the quotient ``Q''  displayed above depends critically on the decision to vaccinate or not made by a very small subset $A$ of residents in England during the study period. This subset $A$ consists of those persons, ages 13 and up,  hospitalized for myocarditis in temporal proximity of a positive COVID-test during the study period while unvaccinated.  As we indicated above, if none of the members of $A$ had later chosen to vaccinate during the study period, there would be no test-positive u-myocarditis cases in the study population, and Patone et al's analysis would have yielded 0 risk to the unvaccinated of hospital admission for test-positive-associated myocarditis.  On the the other hand, suppose $114$ is the total number of persons in $A$; that is, every unvaccinated resident of England, age $\ge 13$, who, during the study period, is hospitalized with test-positive-associated myocarditis later chose to vaccinate during the study period.  Then 114 is the total number, during the study period, of unvaccinated persons in {\it the entire vaccine-eligible general population}  to be hospitalized for test-positive-associated myocarditis.  In this case, the actual incidence of positive-test-associated myocarditis cases for the general population is 
    $$
     \frac{114}{2958026 + U}
     $$
     where $U$ is the  number of initial positive tests, during the study period, among vaccination-age persons in the general population (13 and up) who remained unvaccinated at the end of the study period. Thus, if 114 represents the number of persons in $A$, then the quotient $114/2958026$ on which Patone et al.\ relied to estimate positive-test-associated myocarditis incidence among the unvaccinated may significantly overstate the incidence. 
    
      Let $B$  be the subset of residents of England, ages 13 and up,  who had a positive SARS-CoV-2 test during the study period while unvaccinated.  Note that $A$ is a very small subset of $B$, consisting of those in $B$ who have been hospitalized for myocarditis during the study period within 28 days of a positive test.  The vast majority of persons in $B$ have experienced a mild (possibly asymptomatic) or moderate case of COVID-19.     The positive-test-associated incidence of myocarditis {\it in the general population} ``IGP'' is then
      $$
    \text{IGP} :=  \frac{\text{number of persons in}\ A}{\text{number of persons in}\ B}
      $$
      provided we ignore potential repeated myocarditis hospitalization for members of $A$ as well as repeated positive tests for members of $B$. (Patone et al.\ follow the same practice in their analysis, which is confined to their study population.)  Let $p_A$ be the probability that a person in $A$ chooses to vaccinate and $p_B$ be the probability that a person in $B$ chooses to vaccinate, so that $114 = p_A \times (\text{number of persons in}\ A)$ and $2958026 = p_B \times (\text{number of persons in}\ B)$.   The  quotient $Q$ that Patone at al.\ use  to measure the incidence of positive-test-associated hospitalizations for myocarditis among members of the study population not yet vaccinated  can now be viewed as follows:      $$
     38.54\approx  \frac{114}{2958026} = \frac{p_A \times (\text{number of persons in}\ A)}{p_B \times (\text{number of persons in}\ B} = \frac{p_A}{p_B}\, \frac{\text{number of persons in}\ A}{\text{number of persons in}\ B} =  \frac{p_A}{p_B}\, \text{IGP}.
      $$

       Thus, Patone et al.'s analysis will overstate the incidence of positive-test associated hospitalization for myocarditis in the general population ``IGP'' if $p_A$ exceeds $p_B$.  It's easy to argue this might be the case.  The vast majority of those in $B$ have recovered from a mild to moderate COVID infection, have natural immunity, and perhaps are aware of \href{https://www.mdpi.com/2075-1729/11/3/249/htm}{\textcolor{blue}{\underline{studies}}} showing that COVID-recovered individuals are more likely to experience severe side effects from COVID vaccination than those not previously infected.  In contrast,  those in $A$ have experienced a severe COVID case associated with a hospitalization for myocarditis; their fear of a repeat infection and possible encouragement to vaccinate from doctors they've encountered during their hospitalization might lead to $p_A$'s being quite high, e.g., $0.9$.  If $p_A = 0.9$ and $p_B = 0.6$, then Patone et al.\ have overstated the risk of positive-test-associated myocarditis  by a factor of $0.9/0.6 = 1.5$.   To put this into context, assuming $p_A = 0.9$ and $p_B = 0.6$, then the IRR estimate of the incidence of myocarditis {\it following infection} for men under 40  computed earlier would be further reduced to $\frac{1}{1.5} \times 2.75\approx 1.83$, which, according Table 3, falls below the IRR for all COVID-vaccine doses (including a Pfizer booster) except for a first dose of AstraZeneca ChAdOx1.   Furthermore, the IRR corresponding to the incidence of myocarditis {\it following infection} for  {\it all those under 40}  would fall below the the IRR of 2.59 corresponding a second dose of Pfizer's BNT162b2 and the IRR 2.76 corresponding to a first dose of Moderna's mRNA-1273: $\frac{1}{1.5}\frac{1}{1.58}5.25 \approx 2.22$ is less than both $2.59$ and $2.76$, where  5.25 is the IRR, reported in Table 3 (age $< 40$), corresponding to myocarditis incidence after a positive test (unvaccinated). 
         
  We have demonstrated that Patone et al.'s assessment of risk among the unvaccinated in the general population of  myocarditis hospitalization after a positive COVID test depends on the authors' unjustified assumption that the probabilities $p_A$ and $p_B$ described above are approximately equal and that if $p_A$ is substantially greater than $p_B$, then the risk of myocarditis after {\it COVID infection} in the unvaccinated under age 40 is likely less than that after dose 2 of Pfizer's BNT162b2 and doses 1 and 2 of Moderna's mRNA-1273.
          
     Other potentially unjustified assumptions that    Patone et al.\ make, which relate to study design, are described on page 2 of their {\it Circulation} article as follows:
     \begin{quotation}{\small Any time-varying factors, such as seasonal variation, need to be adjusted for in the analyses. Hospital admissions were likely to be influenced by the pressure on the health systems because of COVID-19, which was not uniform during the pandemic study period. To allow for these underlying seasonal effects, we split the study observation period into weeks and adjusted for week as a factor variable in the statistical models.}
     \end{quotation}
     Patone at al.\ do not provide any details about the adjustments relating to seasonality and hospital-admissions pressure made in their statistical analysis. Thus, readers and reviewers cannot assess if the adjustments implemented are justified.  There is evidence that myocarditis is not seasonal; e,g.,  \href{https://pubmed.ncbi.nlm.nih.gov/31431396/}{\textcolor{blue}{\underline{one study}}} \cite{Skajaa}   ``Lack of seasonality in occurrence of pericarditis, myocarditis, and endocarditis''  concludes, ``The data indicate no important seasonal variation in the occurrence of pericarditis, myocarditis, and endocarditis in Denmark between 1994 and 2016.''
     
     \section{A Last-Minute Change in Study Design \& Missing or Miscategorized Myocarditis-Death Data}
     
       Reading  \href{https://www.medrxiv.org/content/10.1101/2021.12.23.21268276v1}{\textcolor{blue}{\underline{the preprint version}}} \cite{PatonePP} of Patone et al.'s {\it Circulation} article  reveals that, as originally designed, Patone et al's study did not include an analysis of  the incidence of positive-test associated myocarditis among the unvaccinated. Rather, positive-test-associated myocarditis events,  pre-first-dose and post-first-dose, were combined to compute myocarditis incidence following a positive test independent of vaccination status.     Thus, the original study design did not include the flaw discussed in the preceding section.   The bottom line is that  the flaw in Patone et al's study design described in the preceding section was introduced after nearly all study data had been collected and analyzed;  in fact, apparently, the flaw was introduced during {\it Circulation}'s peer- and editorial-review process. 
          
    To illustrate how the analysis of positive-test-associated myocarditis changed from the preprint version of Patone et al.'s {\it Circulation} article to the published version, first consider the following paragraph from the preprint: 
     \begin{quotation}{\small \sl Preprint Version, Paragraph Following Table 1}: {\small In males aged less than 40 years, we observed an increased risk of myocarditis in the 1--28 days following a first dose of BNT162b2 (IRR 1.66, 95\%CI 1.14, 2.41) and mRNA-1273 (IRR 2.34, 95\%CI 1.03, 5.34); after a second dose of ChAdOx1 (2.57, 95\%CI 1.52, 4.35), BNT162b2 (IRR 3.41, 95\%CI 2.44, 4.78) and mRNA-1273 (IRR 16.52, 95\%CI 9.10, 30.00); after a third dose of BNT162b2 (IRR 7.60, 95\%CI 2.44, 4.78); and following a SARS-CoV-2 positive test (IRR 2.02, 95\%CI 1.13, 3.61). }
     \end{quotation}
     There is no comparable paragraph in the published version---one in which, for men under 40,  vaccination-associated myocarditis is compared to positive-test associated myocarditis.  However, the portion of Patone et al.'s Table 3 reproduced in Section 2 above makes the comparison.   We summarize below the information in Table 3 relating to men under 40:
     \begin{quotation}{ \sl Published Version,  Table 3}: {\small  In males aged less than 40 years, there was an increased risk of myocarditis in the 1--28 days following a first dose of BNT162b2 (IRR 1.85, 95\%CI 1.30, 2.62) and mRNA-1273 (IRR 3.08, 95\%CI 1.33, 7.03); after a second dose of ChAdOx1 (2.73, 95\%CI 1.62, 4.60), BNT162b2 (IRR 3.08, 95\%CI 2.24, 4.24) and mRNA-1273 (IRR 16.83, 95\%CI 9.11, 31.11); after a third dose of BNT162b2 (IRR 2.28, 95\%CI 0.77, 6.80); and following a SARS-CoV-2 positive test:  (IRR 4.35, 95\%CI 2.31, 8.21) {\bf before vaccination};  (IRR 0.39, 95\%CI 0.09, 1.60) {\bf vaccinated}. }
     \end{quotation}
    Remark: Recall from the discussion of Sections 2 and 3 above that the IRR for {\it infection-associated} myocarditis for the unvaccinated is highly likely to be less than 2.75 and possibly less than 1.83.  
    
    We now provide a dramatic illustration of the incompatibility of the structure of Patone et al.'s study with an assessment of incidence of positive-test-associated myocarditis for the unvaccinated (in a study-population consisting only of vaccinated persons).  We focus on missing or miscategorized data on positive-test-associated myocarditis deaths.   
        
           One of the myocarditis events tracked in the study is death with ``death recorded on the death certificate with the International Classification of Diseases, Tenth Revision code (Table S1) related to myocarditis.''  For death by myocarditis, the event date is the date of death. A person joins the study population only after vaccination, and the person must be alive to vaccinate; so,  any person having a record of a positive-COVID test pre-first-dose who joins the population through vaccination will not have a myocarditis death associated with the pre-jab positive test. Thus, if a study-population member dies from myocarditis, the death will be associated with a vaccination (if within 28 days of the jab), a post-vaccination positive-test (if within 28 days of the test), or just becomes a baseline myocarditis death.  Thus, the only positive-test-associated myocarditis deaths in the study population occur after a breakthrough infection. 
         Let's examine myocarditis-death data appearing in Table 2 in Patone et al's article published in {\it Circulation}. The description of table contents promises to include data relating to ``SARS-CoV-2 Infection''.  
         \begin{center}
     \includegraphics[height=1.6in]{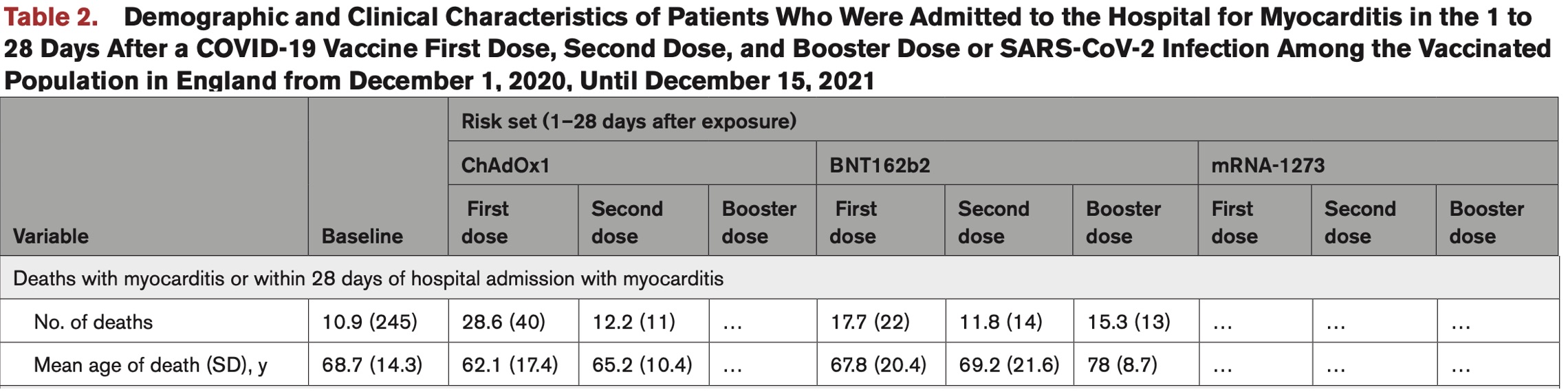}
     
     {\small \sl An excerpt from Table 2 from Patone et al.'s {\it Circulation} article  }    \end{center}
         
    \noindent     If the preceding table does, as the table-header suggests,  provide data on  ``Deaths with myocarditis'' associated with ``SARS-CoV-2 Infection'', where are such deaths recorded?  One possibility is that these deaths are in the baseline column (and account for some of the 245 baseline deaths), but that would be a miscategorization, equivalently, {\it a misrepresentation of fact}.  We suspect the data is simply omitted. Why?  If it were included, then it would be obvious that Patone et al.'s separate analysis of positive-test associated  myocarditis events pre-first-dose vs.\ post-first-dose  is  incompatible with the structure of their study.        
       
        Consider the following excerpt from Supplementary Table 2 of the preprint version \cite{PatonePP} of Patone et al.'s {\it Circulation} article. 
          
             \begin{center}
     \includegraphics[height=1.2in]{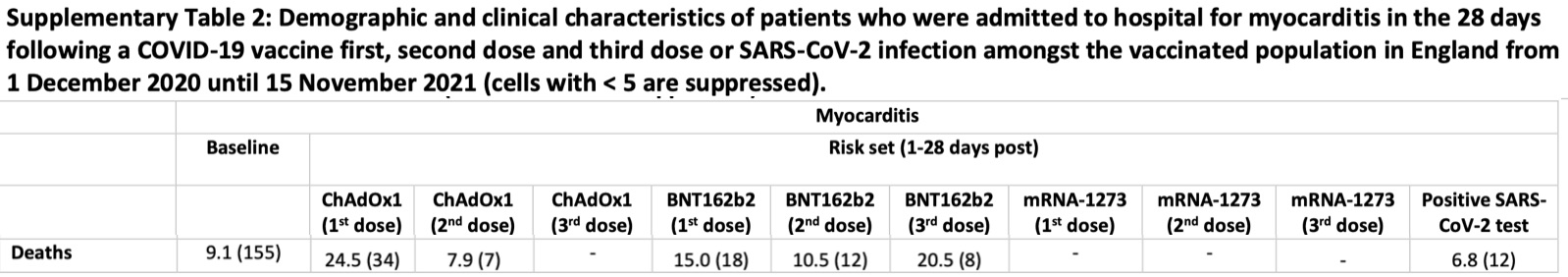}
     \end{center}
        We see that there are 12 positive-test-associated deaths in the study population during the period 1 December 2021--15 November 2021, so that there are necessarily $\ge 12$ positive-test-associated deaths in the study population during the full study-period 1 December 2021--15 December 2021 of Patone et al.'s published article.  As discussed above, the structure of Patone et al.'s study is such that  all positive-test associated myocarditis deaths must occur post vaccination.  Thus, given the way Patone et al.\ chose to analyze positive-test-associated myocarditis for their published study and assuming that positive-test-associated myocarditis deaths are not inappropriately included as baseline deaths, a table providing a complete report of the death-by-myocarditis study outcome would include a number-of-deaths row having the form illustrated below.  .  
        \begin{center}
     \includegraphics[height=1.45in]{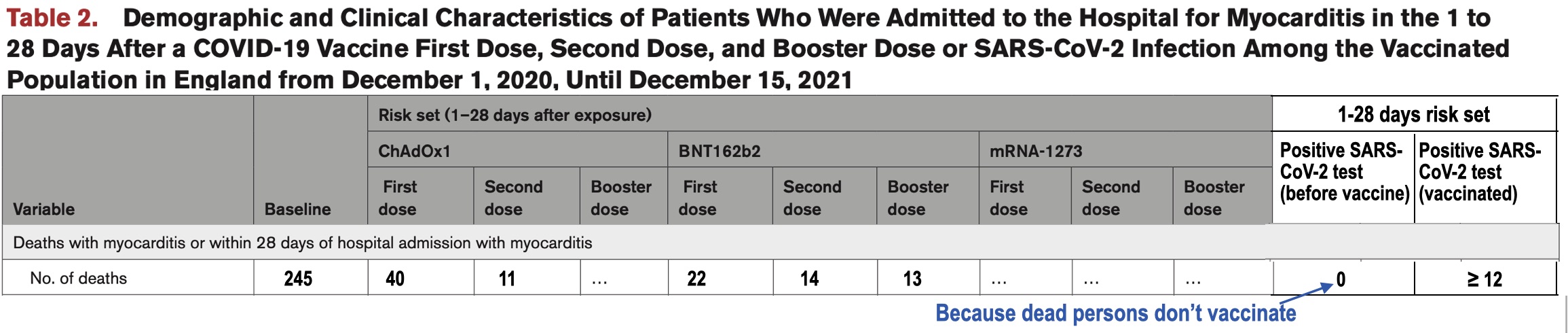}
     \end{center}
    A table like the preceding one was not included in Patone et al.'s article because it clearly shows how inconsistent the structure their study is with the authors' analysis, {\it included in their published Circulation article}, of the incidence of positive-test-associated myocarditis for the unvaccinated in a study-population consisting only of vaccinated persons.  Why did Patone et al.\ make the decision to modify their study design to include such an analysis (apparently while their {\it Circulation} submission was under review for publication)?  
    
     
     \section{Additional Limitations of Patone et al.'s Study}
            Researchers should be especially careful in describing limitations of their findings when their studies might influence public-health policy.  Consider the first two sentences of the ``Discussion'' section of Patone et al's {\it Circulation} article, which read as follows:
            \begin{quotation}{\small In a population of $>42$ million vaccinated individuals, we report several new findings that could influence public health policy on COVID-19 vaccination. First, the risk of myocarditis is substantially higher after SARS-CoV-2 infection in unvaccinated individuals than the increase in risk observed after a first dose of ChAdOx1nCoV-19 vaccine, and a first, second, or booster dose of BNT162b2 vaccine.}
            \end{quotation}

   The preceding statement requires appropriate qualification!
 \begin{itemize}
 \item As we have already noted, the use of ``infection'' is misleading; furthermore, if it's given its obvious meaning, then, as we have demonstrated above, for men under 40, the risk of myocarditis after {\it COVID infection} in the unvaccinated is likely less than that after dose 2 of Pfizer's BNT162b2 (as well as doses 1 and 2 of Moderna's mRNA-1273). 
 \smallskip
 
 \item  The Alpha variant of SARS-CoV-2 was the dominant variant in the UK between December 2020 and May 2021, Delta took over as the dominant variant in May 2021, and Omicron became dominant in December 2021. (See, e.g., the {\it Lancet}  \href{https://www.thelancet.com/journals/laninf/article/PIIS1473-3099(22)00001-9/fulltext}{\textcolor{blue}{\underline{article}}} ``Generation time of the alpha and delta SARS-CoV-2 variants: an epidemiological analysis.'')  According to the UK Health Security Agency's   \href{https://assets.publishing.service.gov.uk/government/uploads/system/uploads/attachment_data/file/1041833/20211216_OS_Daily_Omicron_Overview.pdf}{\textcolor{blue}{\underline{``Omicron daily overview: 16 December 2021''}}},  the total number of confirmed Omicron cases in England, through 6 PM on 15 December 2021  stands at 10,740.   However, there were many unconfirmed cases.     In Section 2 of the appended supplement, we show that a model developed by the UK Health Security Agency suggests that fewer than 1\%  of the 5,934,153 (first) positive COVID tests that contributed to the study's findings indicated Omicron infections.  Clearly, Patone et al.'s risk estimates for positive-test-associated myocarditis among the unvaccinated or vaccinated do not necessarily apply to the Omicron variant, which is the variant of current public-health concern.   
   
   Omicron infection is recognized to be milder than that of previous variants.   \href{https://pubmed.ncbi.nlm.nih.gov/35675841/}{\textcolor{blue}{\underline{A study}}} by Lewnard et al.\ \cite{LEA} suggests reduced hazard ratios for severe clinical outcomes across the board for Omicron versus Delta, with hazard reduction ``starkest among individuals not previously vaccinated against COVID-19''; e,g., the adjusted hazard ratio for mortality is 0.14 (0.07, 0.28) for the unvaccinated.  There is every reason to expect the infection-associated myocarditis hazard ratio---especially in the unvaccinated---is substantially reduced as well. Patone et al.\ do not acknowledge that their findings may not continue to be valid for the Omicron variant.
 \smallskip

 \item  Suppose you are a parent of a child, say between 12 and 15, and you read the two sentences from Patone et al's ``Discussion'' section quoted above. Wouldn't you conclude that your child's risk of myocarditis is ``substantially higher'' after SARS-CoV-2 infection than after vaccination with, e.g., Pfizer's BNT162b2?  Data in eTable 7 from \href{https://jamanetwork.com/journals/jamacardiology/fullarticle/2791253}{\textcolor{blue}{\underline{a study}}} by Karlstad et al.\ \cite{KEA}  in the {\it Journal of the American Medical Association} shows 0 cases of myocarditis following SARS-CoV-2 infection for males and females in the age range 12--15.\footnote{The  study population in the 12--15 age range ``at start of followup'' stands at 1,238,004 and at the end of the followup period 750,253 were unvaccinated.} Thus, it's clear that Patone et al's assertion that myocarditis risk following infection is substantially higher than the increased risk after Pfizer vaccination is quite unlikely to be true for children in the age range 12--15.   In a discussion of limitations of their study in the penultimate paragraph of their article, Patone et al.\ do state the following:
 \begin{quotation}{\small
 [A]lthough we were able to include 2 230 058 children age 13 to 17 years in this analysis, the number of myocarditis events was small (56 events in all periods and 16 events in the 1 to 28 days after vaccination) in this subpopulation and precluded a separate evaluation of risk.}
\end{quotation}
 Because the authors do not mention any positive-test associated events in their 13--17 age group, it appears that data from their study is consistent with that from Karlstad et al.'s, so that both studies combine to suggest that the risk of myocarditis for those 12--17 is highly likely to be  lower following infection than following vaccination.   \smallskip
 
\item The authors might have cited studies reaching conclusions contradicting theirs, e.g.,    “The Incidence of Myocarditis and Pericarditis in Post COVID-19 Unvaccinated Patients-A Large Population-Based Study" by  \href{https://pubmed.ncbi.nlm.nih.gov/35456309/}{\textcolor{blue}{\underline{Tuvali et al.}}} \cite{TEA}, published in the {\it Journal of  Clinical Medicine} 15 April 2022, whose abstract concludes, ``We did not observe an increased incidence of neither pericarditis nor myocarditis in adult patients recovering from COVID-19 infection.''
\smallskip

\item Finally, Patone at al.\ should have noted that their definition of a case of myocarditis among members of the study population is quite restrictive and that other studies have found a much higher incidence of myocarditis and pericarditis post vaccination by counting outpatient diagnoses as well as inpatient diagnoses. For  instance,  ``Risk of myopericarditis following COVID-19 mRNA vaccination in a large integrated health system: A comparison of completeness and timeliness of two methods'' by \href{https://pubmed.ncbi.nlm.nih.gov/35404496/}{\textcolor{blue}{\underline{Sharff et al.}}} \cite{SEA}, e-published  16 April 2022, finds ``Among those who received a second dose of vaccine ($n = 146,785$), we estimated a risk as 95.4 cases of myopericarditis per million second doses administered (95\% CI, 52.1-160.0).''  The post dose-2 estimated risk for men 18--24 is 537.1 cases per million---see Figure 1.  The study by \href{https://jamanetwork.com/journals/jamacardiology/fullarticle/2791253}{\textcolor{blue}{\underline{Karlstad et al.}}}, cited earlier, {which requires a case of myocarditis to be associated with an inpatient hospital admission based on primary or secondary discharge diagnosis for myocarditis, found,  for men 16--24,}  an excess myocarditis case rate of 55.5 per million following a second dose of BNT162b2 (see Table 2), dropping to 5.9 per million for men 25--39, while Patone et al., Table 4, find a post-dose-2 BNT162b2 excess case rate of 11 cases per million for men under 40.  
\end{itemize}
\smallskip

     Before concluding, we would like emphasize that comparing the risk of myocarditis following SARS-CoV-2 infection to the risk following individual doses of vaccines provides an incomplete assessment of risks.  ``Vaccination'' with an mRNA vaccine necessary includes risks associated with two doses, and likely booster doses. Thus, myocarditis risk following infection should be compared to the combined risk of at least doses 1 and 2 of an mRNA vaccine. Of course, there are risks unrelated to myocarditis associated with both infection and vaccination. 
     Finally,  a comparison of a risk associated with infection to the same risk associated with vaccination should not be restricted  to only the 28 days following infection or vaccination.  If vaccination prevented infection and repetition of vaccination weren't required, then limiting risk assessment of infection versus vaccination to a short window during which adverse outcomes typically occur seems reasonable,   However, in the long run, COVID vaccination provides little or no protection from infection.\footnote{According to Table 3 ``Consensus vaccine effectiveness estimates'' from the \href{https://www.gov.uk/government/publications/covid-19-vaccine-weekly-surveillance-reports}{\textcolor{blue}{\underline{4 August 2022 UK Health Surveillance Report}}}: Dose 2 of BNT162b2, provides after 6 months, 20\% (10-30) effectiveness in preventing all infection and 15\% (10-15) against symptomatic infection. A booster dose of BNT162b2, provides after 6 months, 0\% (0-10) effectiveness in preventing all infection and 10\% (0-20) against symptomatic infection. Dose 2 of mRNA-1273, provides after 6 months, 30\% (10-50) effectiveness in preventing all infection and 15\% (10-20) against symptomatic infection.  A booster dose of mRNA-1273, provides after 6 months, 0\% (0-10) effectiveness in preventing all infection and 10\% (0-20) against symptomatic infection.}
 Thus, an analysis of risks versus benefits of vaccination must assess to what extent will vaccination reduce the number of infections a vaccinated person will experience and to what extent, if any, will vaccination reduce the incidence and/or severity of adverse outcomes associated with infections. 
     
    \section{Conclusion}
          Patone et al's  definition of SARS-CoV-2 infection is not reasonable in the context of their study, resulting in exaggerated myocarditis incidence rates associated with  infection.   We have shown that Patone et al.'s assessment of the risk of myocarditis among the unvaccinated in the general population after a positive COVID test depends on the unjustified assumption that the probabilities $p_A$ and $p_B$ discussed in Section 3 above are approximately equal; if $p_A$ is larger than $p_B$, then the result is further exaggeration of myocarditis incidence rates among the unvaccinated associated with SARS-CoV-2 infection.  Because of the unexamined relationship between $p_A$ and $p_B$ as well as the underascertainment of COVID infections occurring in members of the study population while unvaccinated, Patone and et al.\ have not established that for those unvaccinated under 40 the risk  of myocarditis is greater after SARS-CoV-2 infection than after COVID-19 vaccination.   Moreover, in Section 4, we showed that the authors' decision, apparently made while their {\it Circulation} submission was under review for publication,  to provide an assessment of myocarditis risk for the unvaccinated is inconsistent with the structure of their study and resulted in a failure to report or to properly categorize  positive-test associated-myocarditis deaths in their study population.  Finally,  Patone et al.'s study should not influence public-heath policy not only because of the flaws in its methods and assumptions,  but also because of the limitations of their study discussed in Section 5, in particular its lack of meaningful data on myocarditis risk after Omicron infection. 
         \section{Author Contributions}
        PB analyzed the data and drafted the manuscript; SPP edited the manuscript.
       \section{Conflict of interest}
        PB has no relevant conflicts of interest to report. SPP holds a short position on Moderna stock. 
     \section{Acknowledgments}
        No funding was provided for this report. 
        
        \begin{center}
{\Large  \bf Supplement}\
\end{center}
 
\section*{{\small 1. Estimating the Number of SARS-CoV-2 Infections in Members of Patone et al.'s Study Population Before Vaccination}}
Our goal is  to compute a lower bound on the number of SARS-CoV-2 infections occurring during Patone et al.'s study period (1 Dec 2020 -- 15 Dec 2021) in members of the study population while they were unvaccinated.   Equivalently, we compute a lower bound on the number of residents of England, ages 13 and up, receiving, during the study period, a first dose of  a SARS-CoV-2 vaccine after having been infected by COVID-19 during the study period.   We use two sources of data to accomplish this---cumulative percentage estimates of the number of persons in England infected with SARS-CoV-2 from 30 November 2020 through 14 December 2021 from the UK's Office of National Statistics (ONS) and reports of the total number of 1st doses of COVID vaccines administered in England through a given date from the UK's National Health Service (NHS).

   The UK's ONS has made publicly available \href{https://www.ons.gov.uk/peoplepopulationandcommunity/healthandsocialcare/conditionsanddiseases/articles/coronaviruscovid19infectionsurveytechnicalarticlecumulativeincidenceofthenumberofpeoplewhohavetestedpositiveforcovid19uk/22april2022
}{\textcolor{blue}{\underline{a technical article}}}   ``Coronavirus (COVID-19) Infection Survey technical article: Cumulative incidence of the number of people who have tested positive for COVID-19, UK''  that ``presents modelled estimates of the number of people who have had at least one episode of coronavirus (COVID-19) since the start of the UK Coronavirus Infection Survey (CIS) on 27 April 2020 until 11  February 2022.''  The modeling process entails the estimation of ``the daily proportion of the population who would test positive with their first known COVID-19 infection (if they were tested),'' while positive tests are used to predict infections. 

Estimates of the percentage of those in the population who have had at least one episode of coronavirus infection are provided in an {\it Excel} spreadsheet, available for download under item 4  at \href{https://www.ons.gov.uk/peoplepopulationandcommunity/healthandsocialcare/conditionsanddiseases/articles/coronaviruscovid19infectionsurveytechnicalarticlecumulativeincidenceofthenumberofpeoplewhohavetestedpositiveforcovid19uk/22april2022#estimates-of-cumulative-incidence-by-country}{\textcolor{blue}{\underline{the ONS webpage}}}  for the technical article described above.  The data of interest are those used to create Figure 1 (for England) under item 4.  Spreadsheet data are provided to seven decimal places, with display initially limited to 2 decimal places; we chose to compute our underestimate of those infected with COVID before joining the study population (via receiving an initial dose of a COVID vaccine) using the the full seven-decimal-place data in the spreadsheet.    Here's a portion of the spreadsheet data.
\begin{center}
\includegraphics[height = 3.5in]{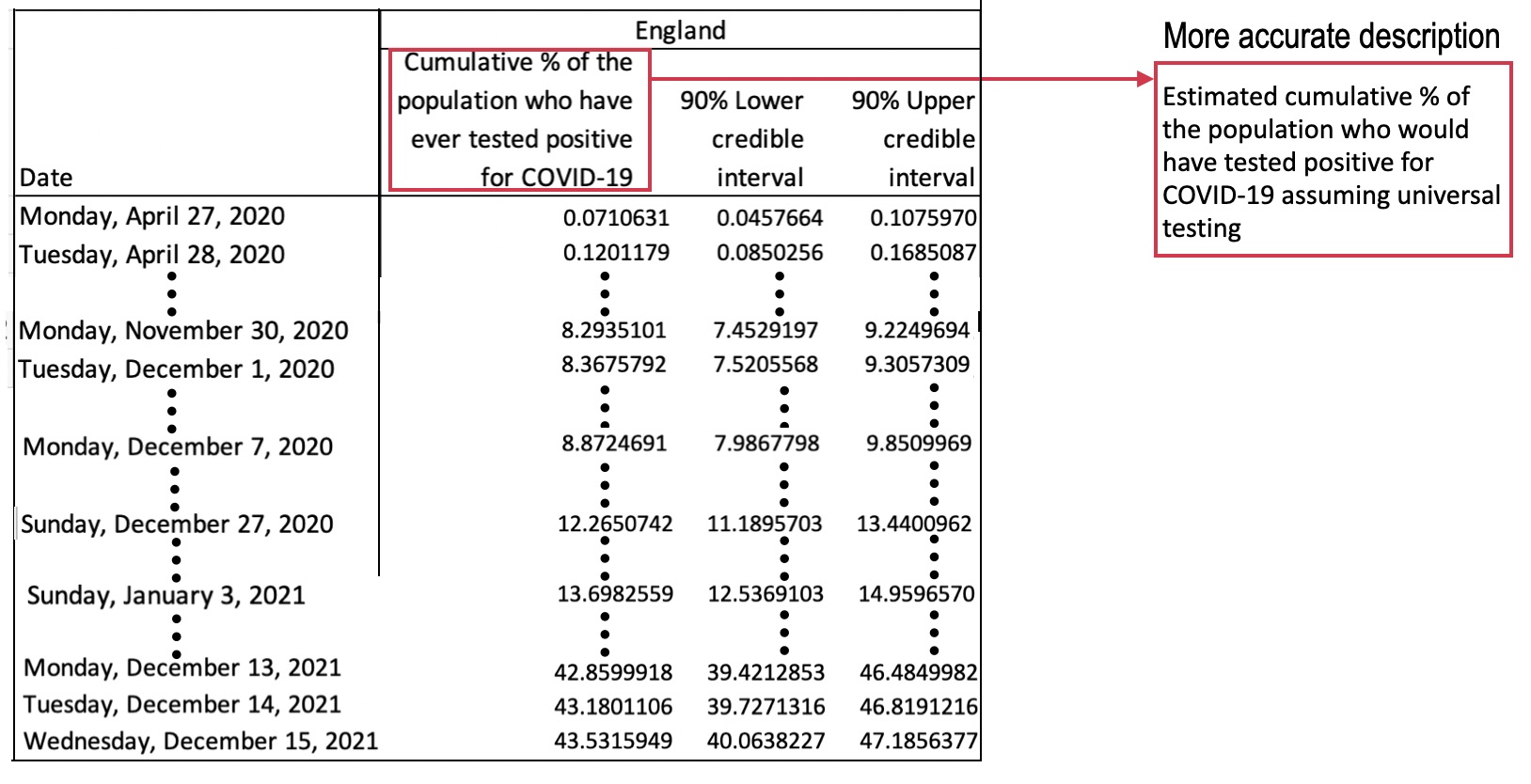}\nopagebreak
\vspace{-.1in}

{\bf Table 1}:  { \sl Some data from an ONS SARS-CoV-2 infection-model spreadsheet}  
\end{center}
We see that on the first day of Patone et al.'s study period (1 December 2020), the spreadsheet provides an estimated increase of 0.0740691\% in the number of persons infected by COVID in England and on the next-to-last day of the study period (14 December 2021), the corresponding increase is 0.3201188\%.\footnote{Patone et al.'s study ignores events the day of a positive test owing to ``small numbers'' (Table-3 footnote); thus, infections occurring 15 December 2021 are irrelevant to the study.}   

    On the first day of the study period, 1 December 2020, we assume that all of the 42,842,345 residents of England who were to become members of Patone et al.'s study population, by receiving at least one dose of a SARS-CoV-2 vaccine, were unvaccinated.  Because England's COVID vaccination program  for the general public began 8 December 2020, we assume that for the first week of the study period all persons who eventually joined the study population remained unvaccinated.   According to data displayed in Table 1 above, the percentage increase in the number of COVID-infected persons in England 1 December 2020 through 7 December 2020 is $(8.8724691-8.2935101)\% = 0.578959\%$.  Thus, we estimate that during the first week of the study period $0.00578959\times 42,842,345 \approx 248,040$ persons in the study population were infected with COVID-19 (infected persons who would later vaccinate for COVID-19).  
    
    To obtain additional estimates of those who were infected before receiving a 1st dose of a COVID vaccine, we need to track numbers of 1st vaccine doses administered during the study period.  At the  NHS's  ``\href{https://www.england.nhs.uk/statistics/statistical-work-areas/covid-19-vaccinations/covid-19-vaccinations-archive/}{\textcolor{blue}{\underline{COVID-19 Vaccinations Archive}}}'' page, there is a ``Weekly Covid-19 vaccinations data archive.''  The first report available ``COVID-19 weekly announced vaccinations 31 December 2020'' provides the total number of persons in England receiving a 1st dose during the period between 8 December 2020 and 27 December 2020, including the endpoint dates.   This is how the data is presented 
    \begin{center}
\includegraphics[height = 1.5in]{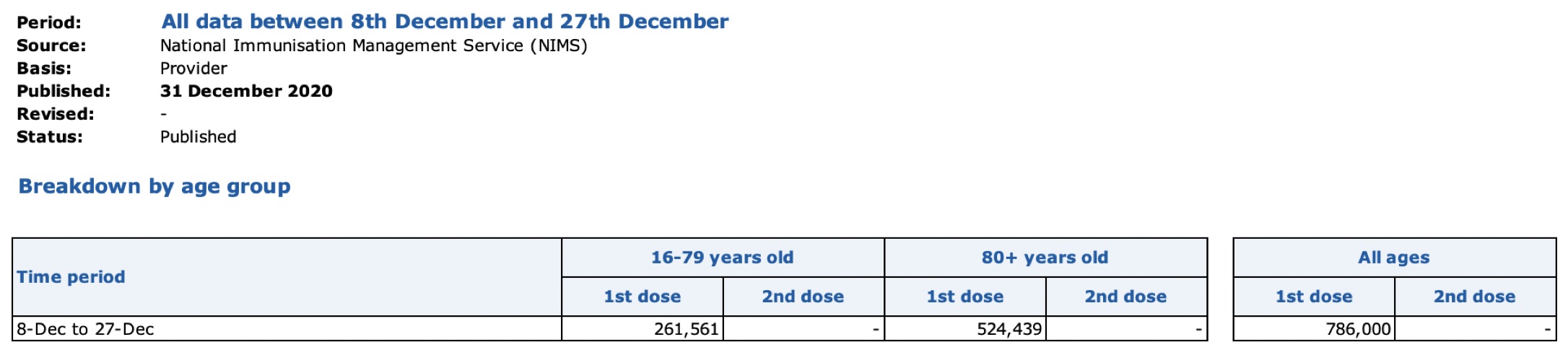}
\end{center}
    The data is also presented  \href{https://www.england.nhs.uk/statistics/wp-content/uploads/sites/2/2020/12/COVID-19-total-announced-vaccinations-31-December-2020.pdf}{\textcolor{blue}{\underline{in narrative form}}} as a document in the listing ``Weekly Covid-19 vaccinations statistical bulletin archive'':

     \begin{center}
\includegraphics[height = 0.75in]{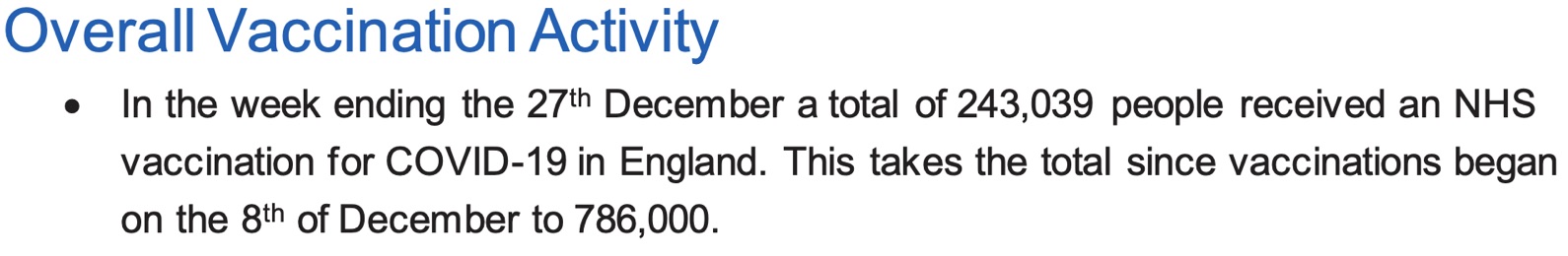}
\end{center}
    The preceding makes it clearer that ``between'' includes the endpoint dates.  

To get a lower bound on the number of persons in the study population who were COVID infected while unvaccinated during the period Dec 8--Dec 27, let's assume that throughout the Dec 8-Dec 27 interval, the number of unvaccinated yet to join the study population is the number on Dec 27 (midnight):
$$
42,842,345 - 786,000 = 42,056,345.
$$
According to the ONS-spreadsheet data appearing in Table 1, the percentage of persons infected increased from 8.8724691\% on Dec 7 to 12.2650742\% on Dec 27, a difference of 3.3926051\%, and thus a lower bound on the number of those eventually joining the study population having contracted, during the period Dec 8--Dec 27, a COVID infection is  
$$
42,056,345 \times 0.033926051 \approx 1,426,806.  
$$
The preceding is a lower bound because during the period Dec 8--Dec 26, the number of unvaccinated yet to join the study population exceeds 42,056,345,  

We continue the process described above, using weekly vaccination reports to obtain a lower bound for each week on the number of unvaccinated yet to join the study population and then multiply by the corresponding percentage increase in the infected population using ONS data illustrated in Table 1.   Here's how this looks in spreadsheet form:
   \begin{center}
\includegraphics[height = 3.4in]{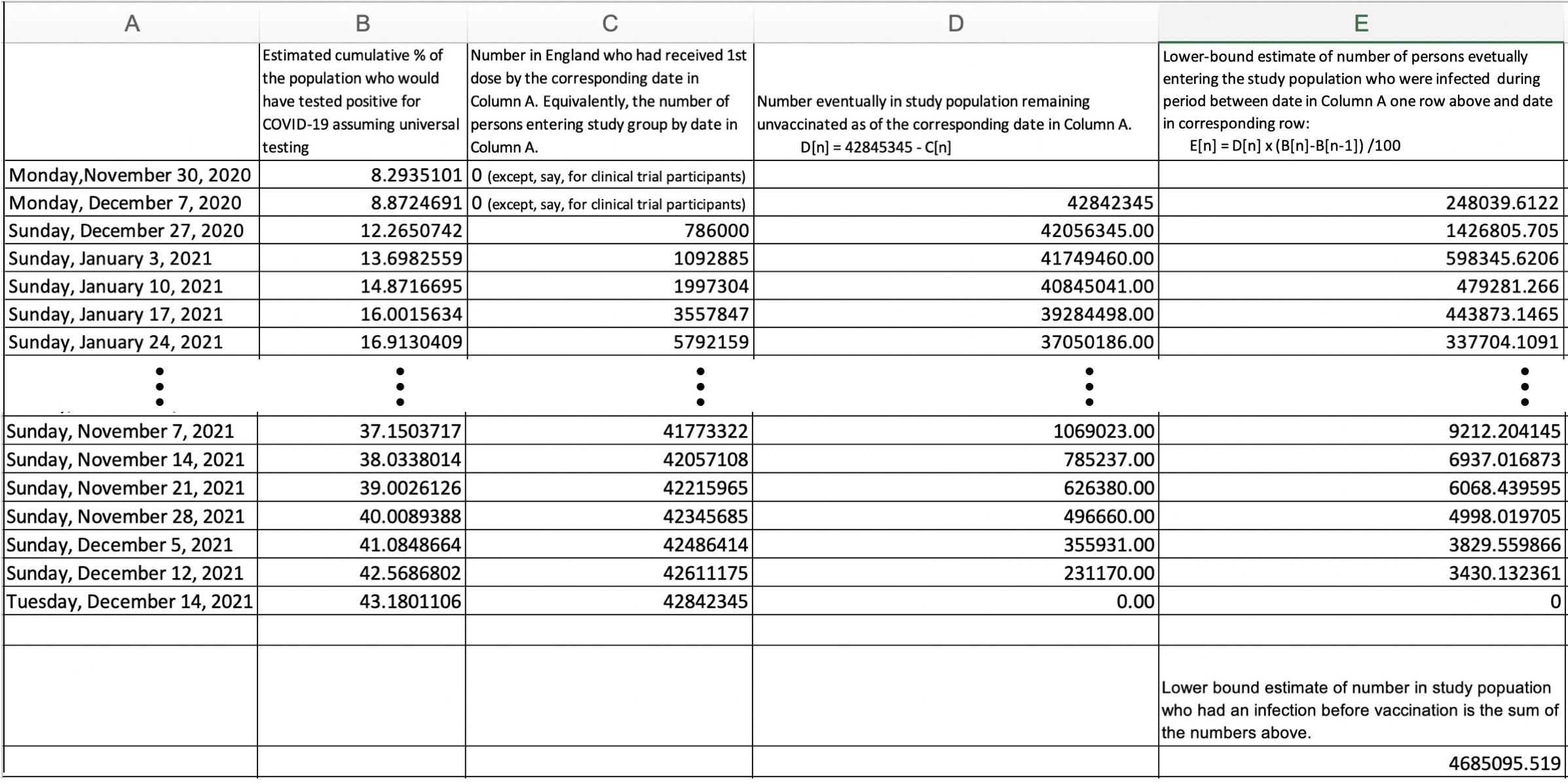}\\
{\bf Table 2}:  {\sl Computation of a lower bound on number infected before first dose of a SARS-CoV-2 vaccine}  
\end{center}
Our lower-bound estimate of the number in Patone et al.'s study population who had a COVID infection during the study period before vaccination is the sum of the numbers in column E---the number  on the lower right of Table 2 above, which we round down to 4,685,095.   

Our justification that the estimate 4,685,095 provides a lower bound for the number of infections that occurred in members of study population during the study period before they were vaccinated is as follows:
\begin{itemize}
\item The computation itself, mathematically, clearly yields a lower estimate, assuming the ONS and NHS data upon which it is based is accurate.

\item The computation ignores repeat infections (which should be counted in the incidence computation). 

\item Consider a percentage change in the infected population of say 1\% over a given week $W$.  This percentage change depends on new infections among unvaccinated persons and new infections among vaccinated persons. Given that vaccination provides some protection from infection, we would expect the number of infections among the unvaccinated during week $W$ to increase more than $1\%$ and among the vaccinated to increase less than 1\%. Thus, using the overall percentage change of 1\% to estimate the number of members of the study population who become infected during $W$ while unvaccinated leads to an underestimate.  We have used such overall percentage changes (from the ONS model) in our computation.
\end{itemize}

\section*{{\small 2. Estimating for the period 27 November 2021 through 14 December 2021 the number of positive COVID-19 Tests in England Reflecting Omicron Infections}}

 In this section, we approximate the number of documented Omicron cases in England that occurred from 27 November 2021 through 14 December 2021.\footnote{Recall Patone et al.'s study ignores events the day of a positive test owing to ``small numbers'' (Table-3 footnote); thus, positive tests reported on 15 December 2021 are irrelevant to the study.} 

 In a document titled ``\href{https://assets.publishing.service.gov.uk/government/uploads/system/uploads/attachment_data/file/1041965/20211216_Omicron_daily_infection_estimation_modelling_updated.pdf}{\textcolor{blue}{\underline{Methodology for estimating daily
infections in England: 16 December 2021}}},'' the UK Heath Security Agency   estimated that ``around 24\% of all COVID-19 positive cases with specimen dates on 11 December in England \ldots were highly likely to be the
Omicron \ldots variant.'' Moreover, the Agency also estimated for the period 27 November through December 13  a constant doubling time for {\it infections} of 1.9 days.  

For the 11th of December, the  \href{https://coronavirus.data.gov.uk/details/cases?areaType=nation&areaName=England}{\textcolor{blue}{\underline{UK's Coronavirus Dashboard}}} reports 40,517 first-episode cases for England.  We assume 24\% of these, 9,724, are Omicron cases. We also assume a doubling time for {\it cases} of 1.9 days. Thus, e.g., we would expect $9,724\cdot 2^{1/1.9}  \approx 14,005$ cases on the 12th, $9,724\cdot 2^{2/1.9}  \approx 20,171$ on the 13th, and $9,724\cdot 2^{-1/1.9}  \approx 6,752$ on the 10th.   Treating December 11 as day zero, so that the 14th of December is day 3 while the 27th of November is day -14, we  obtain approximately 
$$
\sum_{n=-14}^3 9724\cdot 2^{n/1.9} \approx 94,904
$$ 
Omicron cases through December 14th.   These 94,904 cases constitute  approximately 7.73\%  of the total number 1,227,074 of first-episode cases reported for England from 16 November 2021 through 14 December 2021 by the UK Coronavirus Dashboard.   

 Patone et al.'s \href{https://doi.org/10.1101/2021.12.23.21268276}{\textcolor{blue}{\underline{preprint version}}}  of their {\it Circulation} article reports 5,185,772 positive tests among study-population members through 15 November 2021 while their {\it Circulation} article reports 5,934,153 positive tests, a difference of 748,381.  Assuming 7.73\% of these positive tests reflect Omicron infections  provides an estimate of $0.0773\cdot  748,381\approx  57,850$ Omicron cases in the study population, and 57,850 is less than one percent of the number of first positive tests 5,934,153 of members of the study population.

         \end{document}